# *An investigation of GSC 02038-00293, a suspected RS CVn star, using CCD photometry.*

Alastair Bruce, Stewart Cruickshank, Tony Rodda and Mark Salisbury.   *(Members of the 2010 Open University Module S382 Undergraduate Programme).*

## Abstract

We present the results of differential, time series photometry for GSC 02038-00293, a suspected RS CVn binary, using data collected with the Open University's PIRATE robotic telescope located at the Observatori Astronomic de Mallorca between 10 May and 13 June 2010.

A full orbital period cycle in the V band and partial cycle of B and R bands were obtained for GSC 02038-00293 showing an orbital period of 0.4955 +/- 0.0001 days. This period is in close agreement with that of previously published values but significantly different to that found by the All Sky Automated Survey of 0.330973 days.    We suggest GSC 02038-00293 is a short period eclipsing RS CVn star and from our data alone we calculate an ephemeris of JD 2455327.614 + 0.4955(1) x E.   We also find that the previously observed six to eight year cycle of star spot activity which accounts for the behaviour of the secondary minimum is closer to six years and that there is detectable reddening at both minima.

## Introduction

We have undertaken a differential photometric study of GSC 02038-00293 known to display short period modulations in optical magnitude and to emit X-rays with a view to characterize the physical properties of this source.

The target is located at RA 16h 02m 48.22s and Dec +25° 20' 38.2'' in the constellation of Serpens Caput.

The target was identified as an object warranting further investigation by Norton, et al (2007)[1] - with the suggestion it was of the type RS CVn (RS Canum Venaticorum).   That paper presents optical light curves obtained during the SuperWASP[13] photometric survey for 428 periodic variable stars coincident with ROSAT[14] X-ray sources.    (It should be noted that GSC 02038-00293 is known in the Norton [1] paper by its SuperWASP name of 1SWASP J160248.22+252038.2)

RS CVn stars traditionally represent a class of close detached binaries with the more massive primary component being a G-K giant or sub-giant and the secondary a sub-giant or dwarf of spectral classes G to M with a temperature difference of ~1000K, though Pandey (2005)[2] also states that there are no implicit restrictions on the class of the secondary. RS CVn type stars show optical variability (outside eclipses) which is characterised by an amplitude up to 0.6 mag in the V band and interpreted as the rotationally modulated effect of cool spots on their surfaces, a result of increased chromospheric activity.    The X-ray emission is an indicator of an active coronal region believed to be the result of rapid rotation and enhanced magnetic fields.

Fast rotation, Ca II, H and K emission lines and a sub-giant component well within its Roche lobe are usually taken to be necessary features. Short-period RS CVns have a period < 1 day, Hall et al (1976)[3].

At the time of our observations, there are few previous references to GSC 02038-00293 to be found. Bernhard & Frank (2006)[4] and Frank and Bernhard (2007)[6] first suggested it as an eclipsing RS CVn binary with an orbital period of 0.495410 days. This is similar to the period indicated by SuperWASP observations (Norton et al. 2007)[1]. Bernhard & Frank (2006)[4] also estimated a long star spot activity cycle generating a 6 to 8 year cycle.

Several weeks after our observations, Korhonen et al (2010)[16] obtained intermediate Hα and low resolution spectroscopic data showing GSC 02038-00293 to be a mid type K star with $T_{eff}$ 4750 and *v* sin*i* = 90 km s$^{-1}$ indicating a rapidly rotating star.

## Observations

We observed GSC 02038-00293 regularly on 10 nights over a period of 34 days in May and June 2010.

### *Equipment*

At the time of our observations, the Physics Innovations Robotic Astronomical Telescope Explorer (PIRATE) facility comprised a remotely controllable 35 cm, f/10 Schmidt Cassegrain telescope (Celestron-14) equipped with an SBIG STL 1001E CCD camera with 1024x1024 24µm pixels, resulting in a field of view of 21 arcmin and a pixel scale of 1.21 arcsec per pixel. (For reference, in August 2010 the OTA was replaced with a Planewave CDK-17 f6.8 astrograph. Refer to Lucas, RJ and Kolb, U, (2011)[18].

The CCD camera is equipped with an 8 position filter wheel and has 5 broadband filters of which only R, V and B were used for this study.

The OTA is mounted on a Paramount ME, a robotic German Equatorial Mount manufactured by Software Bisque, all housed in a remotely controllable 3.5 m dome manufactured by Baader Planetarium, which is on top of the main observatory building at the Observatori Astronomic de Mallorca (OAM), Longitude E 2° 57' 06''; Latitude N 39° 38' 38''; Altitude 203 m. (See PIRATE website http://pirate.open.ac.uk/).

The main user control interface is the software programme ACP Observatory Control (by DC-3 Dreams) with TheSky driving the Paramount mount. Data reduction and processing, and display were done using MaximDL Version 5.08 from Diffraction Limited.

### *Data Acquisition*

The observers' log is summarized in Table 1. It shows the dates observations were made; the start time of each automated run; the run reference letter allocated to the images; the filter used; the exposure length in seconds; and finally, the number of exposures in the run. Where multiple filters were used in a single automated run the bracketed numbers show the number of exposures per filter. So, 10 x (3,2,3) gives 30 V filter, 20 B filter and 30 R filter.

| Date | Start | Run | Filter | Exposure length (seconds) | Number of exposures in run. |
|---|---|---|---|---|---|
| 10/5/10 | 20:24 | A | V | 30 | 30 |
|  | 20:56 | B | V | 30 | 60 |
|  | 21:52 | C | V | 30 | 3 |
| high humidity forced early dome closure ||||||
| 13/5/10 | 20:13 | A | V | 60 | 60 |
|  | 21:36 | B | V | 60 | 60 |
|  | 23:05 | C | V | 60 | 9 |
|  | 23:22 | D | V | 60 | 19 |
|  | 23:53 | E | V | 60 | 27 |
| increasing haze/cloud ended observations early ||||||
| 16/5/10 | 20:25 | A | V | 30 | 60 |
|  | 21:15 | B | V | 30 | 60 |
|  | 22:08 | C | V | 30 | 60 |
|  | 23:01 | D | V | 30 | 60 |
|  | 23:53 | E | V | 30 | 28 |
|  | 00:38 | F | V | 30 | 100 |
|  | 01:58 | G | V | 30 | 100 |
| slight haze, good conditions ||||||
| 22/5/10 | 21:06 | A | V | 30 | 60 |
|  | 21:57 | B | V | 30 | 120 |
|  | 23:42 | C | V | 30 | 60 |
| good conditions ||||||
| 25/5/10 | 20:22 | A | V | 60 | 6 |
|  | 20:33 | B | V | 30 | 30 |
|  | 20:58 | C | V | 30 | 30 |
|  | 21:36 | * | V,B,R | 30 | 50 each filter |
|  | 23:45 | F | V,B,R | 60 | 20 each filter |
|  | 01:10 | G | V,B,R | 30 | 20 each filter |
|  | 02:00 | D* | V,B,R | 30 | 30 each filter |
| clouds at start, improving towards end, bright Moon ||||||
| 28/5/10 | 20:36 | A-D | V,B,R | 30,60,30 | 30 each filter |
|  | 00:05 | E | V,B,R | 30,60,30 | 30 each filter |
|  | 01:38 | F | V,B,R | 30,60,30 | 30 each filter |
| weather prevented early observations, near full Moon ||||||
| 31/5/10 | 20:32 | A | V,B,R | 30,60,30 | 24 each filter |
|  | 21:42 | B | V,B,R | 30,60,30 | 3x(3,2,3) |
|  | 23:03 | C | V,B,R | 30,60,30 | 10x(3,2,3) |
|  | 00:25 | D | V,B,R | 30,60,30 | 10x(3,2,3) |
|  | 01:46 | E | V,B,R | 30,60,30 | 10x(3,2,3) |
|  | 02:56 | F | V,B,R | 30,60,30 | 5x(3,2,3) |
| later cloud, dome failure during eclipse, bright Moon ||||||
| 3/6/10 | 20:23 | A | V,B,R | 30,60,30 | 10x(3,2,3) |

| | 21:45 | B | V,B,R | 30,60,30 | 10x(3,2,3) |
| --- | --- | --- | --- | --- | --- |
| | 23:05 | C | V,B,R | 30,60,30 | 10x(3,2,3) |
| | 23:43 | D | V,B,R | 30,60,20 | 10x(3,2,3) |
| | 00:50 | E | V,B,R | 30,60,20 | 10x(3,2,3) |
| | 02:04 | F | V,B,R | 30,60,20 | 10x(3,2,3) |
| good conditions, R exposure lowered due to saturation | | | | | |
| | | | | | |
| 13/6/10 | 20:34 | A | V,B,R | 30,60,30 | 10x(3,3,3) |
| | 22:08 | B | V,B,R | 30,60,20 | 1x(3,3,3) |
| | 22:30 | C | V,B,R | 30,60,20 | 10x(3,3,3) |
| | 23:58 | D | V,B,R | 30,60,20 | 10x(3,3,3) |
| | 01:24 | E | V,B,R | 30,60,20 | 7x(3,3,3) |
| good conditions, cloud at end of night | | | | | |

*Table 1: Journal of observations for GSC 02038-00293.*

## Method

The finder chart showing the location of GSC 02038-00293 (shown by its SuperWASP name 1SWASPJ160248.22+252038.2) is annotated with reference and check stars for photometry is reproduced in figure 1.

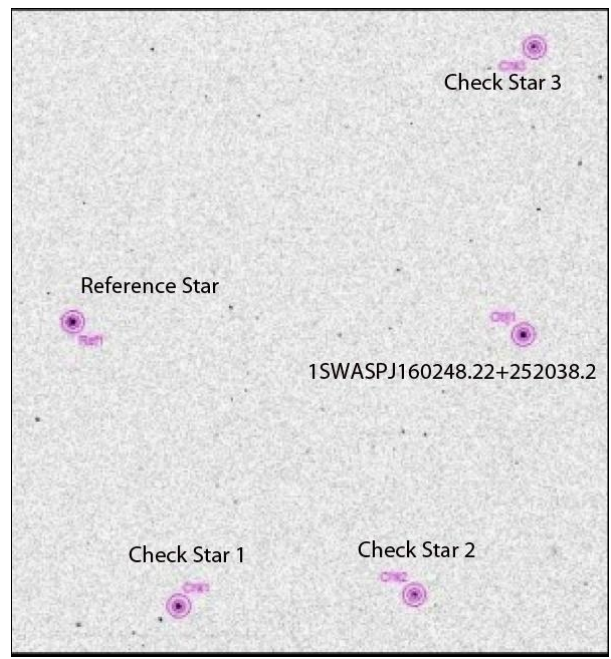

*Figure 1: PIRATE image of* GSC 02038-00293 *with reference and check stars for photometry annotated. Field of view is 15.4'×15'4. North is up and East to the left.*

The field of view was deliberately offset for GSC 02038-00293 in order to provide a more appropriate reference star. The image was centred at 16:03:12.22 +25:20:38.2 (J2000) from 13th May onwards.    It should be noted that for the evening of the 10th May, the reference star lay nearer the edge of the frame.

Table 2 details the reference and check stars used in the analysis of frames from all nights.

|  | Name | RA, Dec (J2000) | V Mag | B Mag |
|---|---|---|---|---|
| Ref Star1 | GSC 2038-0652 | 16 03 38.05 +25 21 10.35 | 9.98±0.3* | 10.694±0.034 |
| Check Star 1 | GSC02038-00987 | 16 03 26.80 +25 13 54.3 | 11.51±0.37‡ | - |
| Check Star 2 | CMC14 J160300.6+251406 | 16 03 00.66 +25 14 06.4 | 13.363* | - |
| Check Star 3 | GSC02038-00648 | 16 02 46.26 +25 22 55.9 | 13.34±0.4‡ | - |

*Table 2: Reference and Check stars for GSC 02038-00293*

\* Data from VizieR using the Carlsberg Meridian Catalogue 14 (CMC14) I/304.

‡ Data from GSC v1.1, photographic magnitudes.

**Data Reduction and Photometry**
All data were bias and dark frame subtracted and then flat fielded with twilight flats using Maxim DL. Aperture settings of 5 pixels, gap 8 pixels, and annulus 3 pixels were used.

# Results

**Light curve fit.**
Using the period analysis software package PERANSO [17] and its Analysis of Variance (ANOVA) fitting method, the best estimate of the orbital period obtained is 0.4955 +/- 0.0001 days.

The following graphs have been phase folded using that orbital period with zero point at JD 2455327.614. Figure 2 shows two complete phases of V band data. Figure 3 shows two phases of data from evenings where observations were performed in B, V and R band.

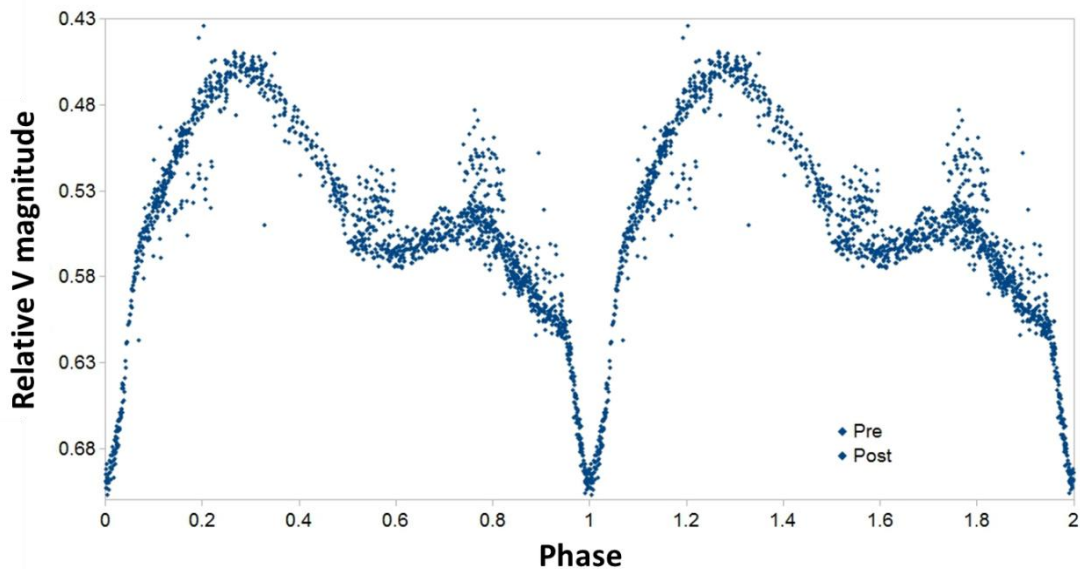

*Figure 2: Relative V band magnitude against phase, two complete cycles.*

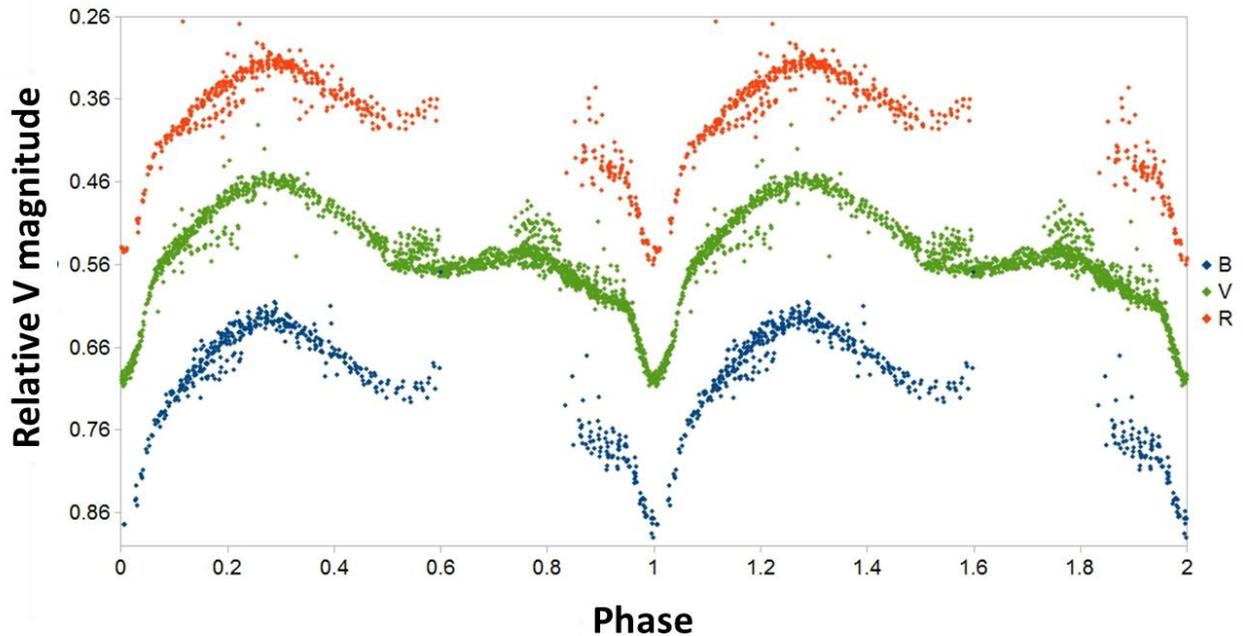

*Figure 3: Relative magnitude against phase, two cycles are shown with R band (top), V band (middle) and B band (bottom)*

It can be seen from both figures 2 and 3 that the primary eclipse, shown at phase 1, is very well defined. The secondary eclipse, suspected at phase 0.5 is less well defined due in part to the increased scatter around phase 0.5-0.8 and the incompleteness of our B and R band curves and the asymmetry of the two maxima.   The primary eclipse itself is slightly asymmetric with indications of a bell-shaped curve at the minimum as opposed to a sharp change.

Both the R and B band curves show broadly the same features as in V.    One notable difference is the sharp upturn in brightness at phase 0.6 following the secondary minimum, most notably in B. There is also a less pronounced shelf coming out of the primary minimum in B.

Figure 4 shows a magnified section of the V band curve between phase 0.4 and 0.8 containing the increased scatter mentioned above.    The data here are displayed according to the date the observations were made to better illustrate the observed changes.    The evenings of the 10$^{th}$ May, 25$^{th}$ May and 13$^{th}$ June show a marked increase in scatter and brightness around phases ~0.55 and ~0.78.   Over the period from the 13$^{th}$ May to 22$^{nd}$ May there seems to be a slight increase in overall brightness, perhaps a prelude to the subsequent high-scatter data on the 25$^{th}$ May but this is unclear.

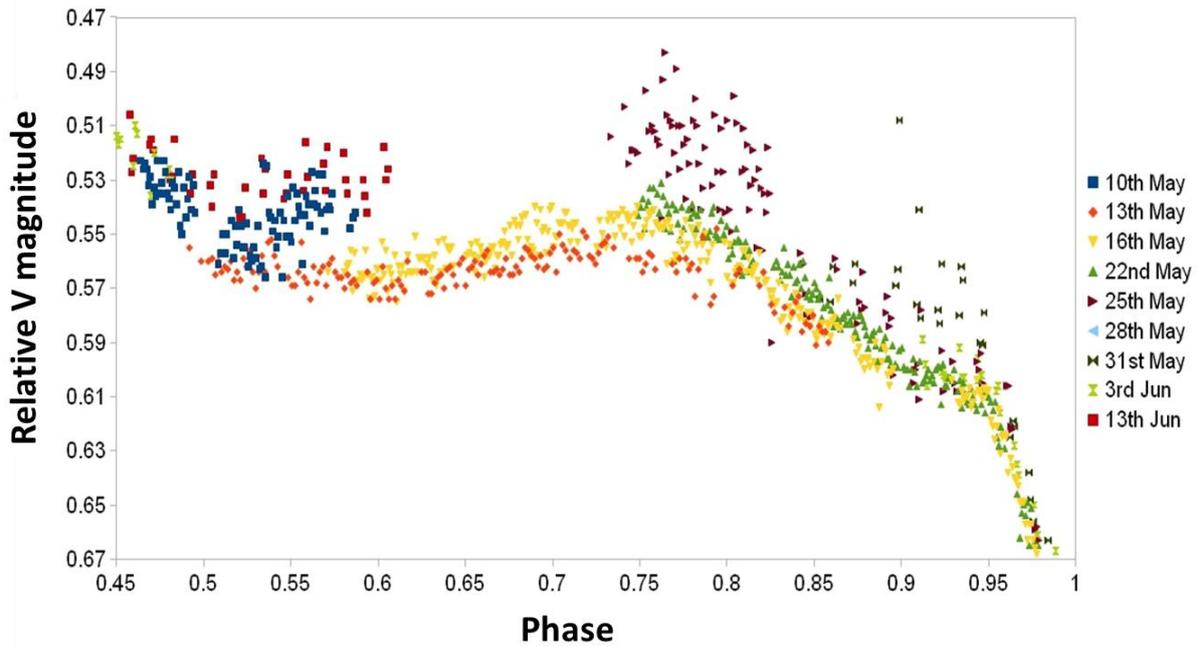

Figure 4 *Relative V band magnitude against phase. Each observing night's data displayed separately*

**B-V changes during eclipse.**

We observed a reddening at the primary minimum, see Table 3. The calculated B-V figures point towards a spectral class G which is in line with RS CVn class systems.    This assumes a negligible contribution from the non-active binary component.    Korhonen et al (2010)[16] obtained spectroscopic data showing GSC 02038-00293 to be a mid type K star.

The spread in our V data meant we were unable to confirm any reddening at the secondary minimum that would indicate star spot activity as observed by Bernhard and Frank (2007)[6] . Pandey (2005)[2] also mentions that reddening outside of the eclipse lends support to the starspot hypothesis.

|  | V | B | B-V |
|---|---|---|---|
| Primary eclipse | 0.70 (10.68) | 0.89 (11.58) | 0.19 (0.90) |
| Primary maximum | 0.46 (10.45) | 0.62 (11.31) | 0.16 (0.86) |

*Table 3: Approximate magnitudes for different parts of the cycle, the numbers in brackets are corrected against our reference star assuming the values B band (10.694) and V band (9.98) obtained from CMC14.*

**Uncertainties and Corrections**

Figure 5 represents a typical observing night's data showing the relative magnitude between the reference star, target and check stars.    The curves have been offset vertically.

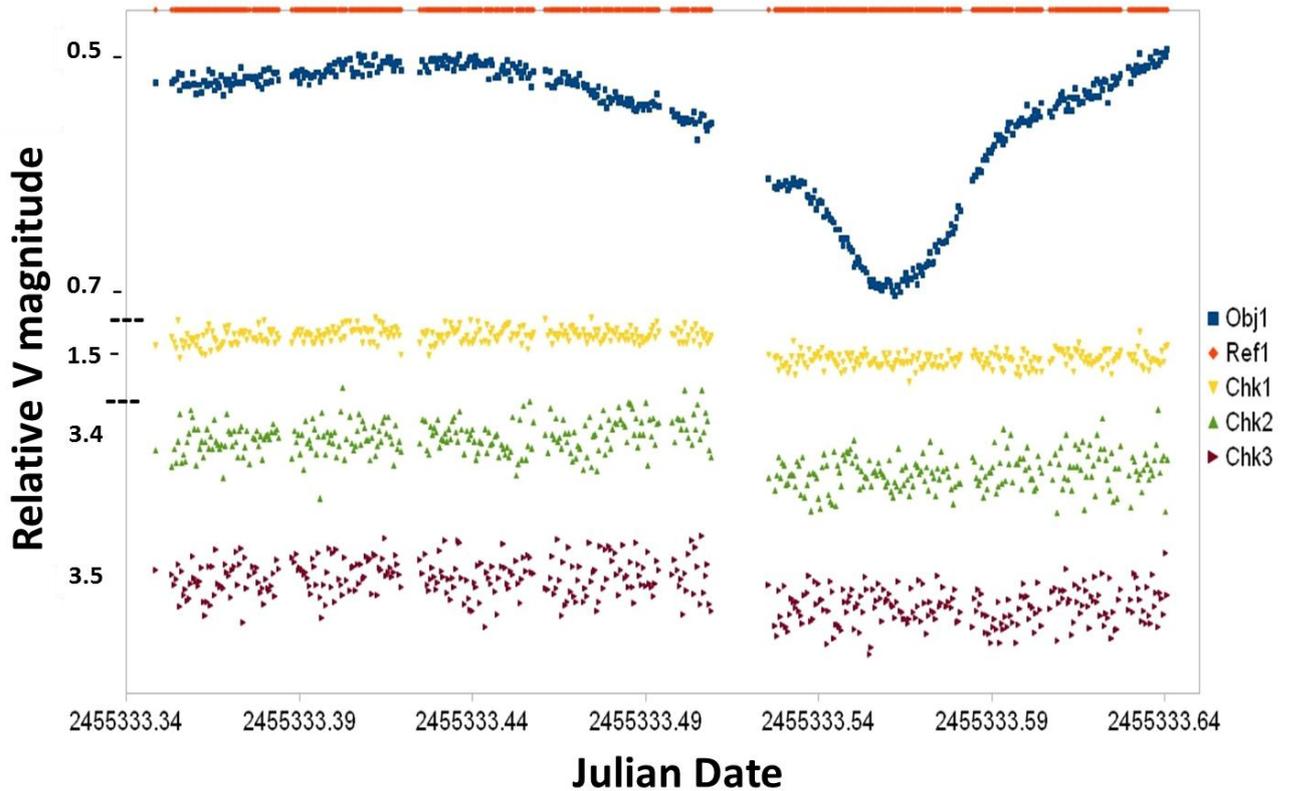

*Figure 5 Reduced V band data obtained from the 16[th] May showing relative V magnitude against JD. NOTE the 'Relative V magnitude axis is compressed to show all data sets.*

Most notable in figure 5 is the step in the data, seen after the telescope underwent a pier flip.

Data collected from the mark one PIRATE facility demonstrated a systematic gradient in the East-West plane that calibration could not remove.   As our target and reference star both lay in this plane the FWHM differences before and after the meridian flip were most accentuated by this effect which manifested itself as a "step" in the pre and post meridian flip data.    (This effect is discussed at length in Holmes, S et al (2011) [(5)].    PIRATE MKII does not exhibit this gradient).

From our study of this effect we chose to apply a uniform post-flip correction to the data for each band: B (-0.055), V (-0.045), R (-0.035).    The confidence of the ANOVA fit in PERANSO was used in determining these values.    Figure 6 shows the uncorrected V data with pre and post-flip data shown separately. The good correlation in the two curves allowed us to be confident in the uniform pier-flip correction.

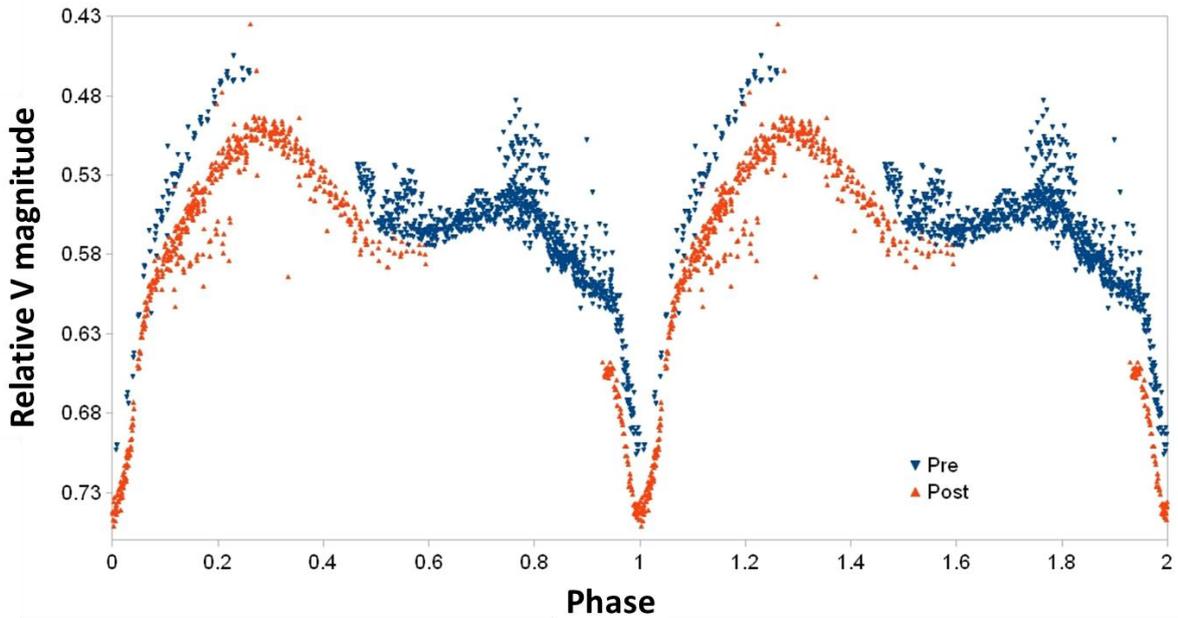

*Figure 6 Uncorrected V band data against phase with pre/post flip data shown separately*

**Photometric uncertainties.**

We have chosen not to show uncertainty bars on the light curve plots for clarity.    Typical photometric uncertainties were calculated as 0.03 magnitude.    An example of such uncertainties applied to our data is provided in Figure 7 below and shows the light curve around minimum on the 16th May. As can be seen, two thirds of the data points lie on, or are touching, the added polynomial trend-line indicating good data quality for this night.

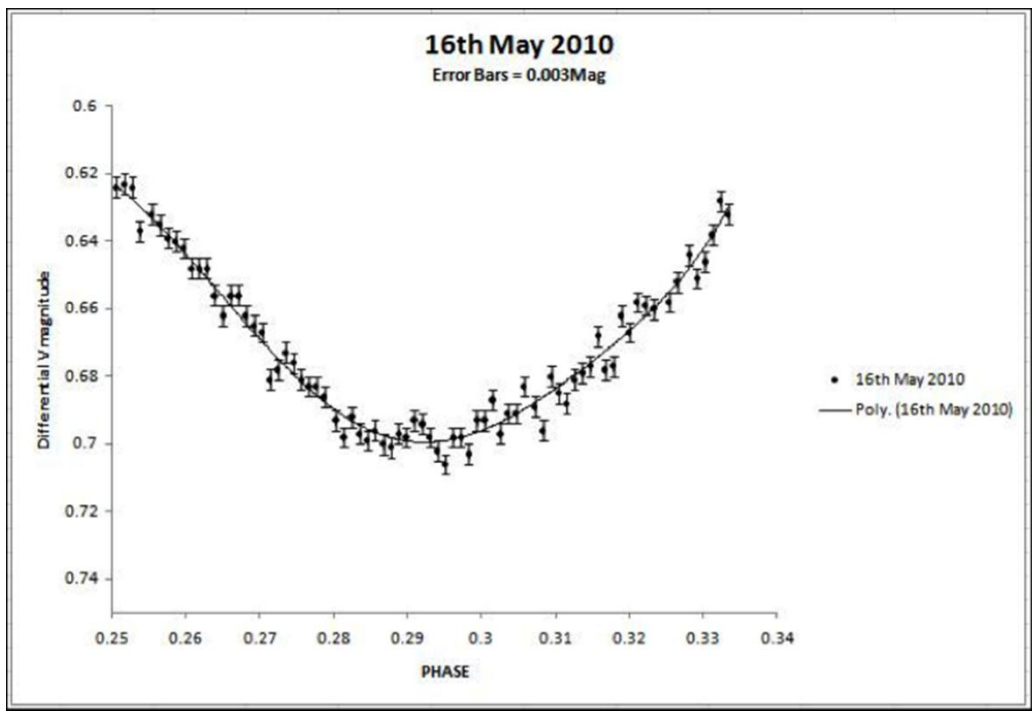

*Figure 7: Error bars of 0.03Mag added to a close up of the primary eclipse minimum (note the small increase in brightness just before minimum).*

Table 4 shows the standard deviation of the check star #1 from the analysis and gives an indication of the seeing on any particular observing night.

| Date | Pre-flip | Post-flip |
|---|---|---|
| 10/05/10 | 0.009 | Not applicable |
| 13/05/10 | 0.006 | Not applicable |
| 16/05/10 | 0.007 | Not applicable |
| 22/05/10 | 0.005 | 0.006 |
| 25/05/10 | 0.017 | 0.007 |
| 28/05/10 | Not applicable | 0.007 |
| 31/05/10 | 0.021 | 0.020 |
| 03/06/10 | 0.006 | 0.007 |
| 13/06/10 | 0.008 | 0.008 |

*Table 4: The standard deviation seen in the check star 1 magnitude for each observing session as an indication of photometric uncertainty in the data*

## Discussion

Our period is calculated as 0.4955 +/- 0.0001 days, significantly different from the All Sky Automated Survey [15] period of 0.330973 days. The ASAS data is noisy and less well sampled and may be one reason for the discrepancy.

Bernhard and Frank (2006 – 2010)[4][6] have studied the light-curve of GSC 2038-0293 since 1999. They have observed that while the magnitude of the primary minimum stays roughly the same, the magnitude of the secondary minimum varies from year to year and is believed to be caused by star spot activity. Our observations support these findings with a light curve that fits both in general shape and a secondary-minimum cycle in close agreement to the proposed 6 to 8 year star spot cycle, see Figure 8 below where we have superimposed our data (for 2010) onto existing 1SWASP data from years 2004, 2007 and 2008.

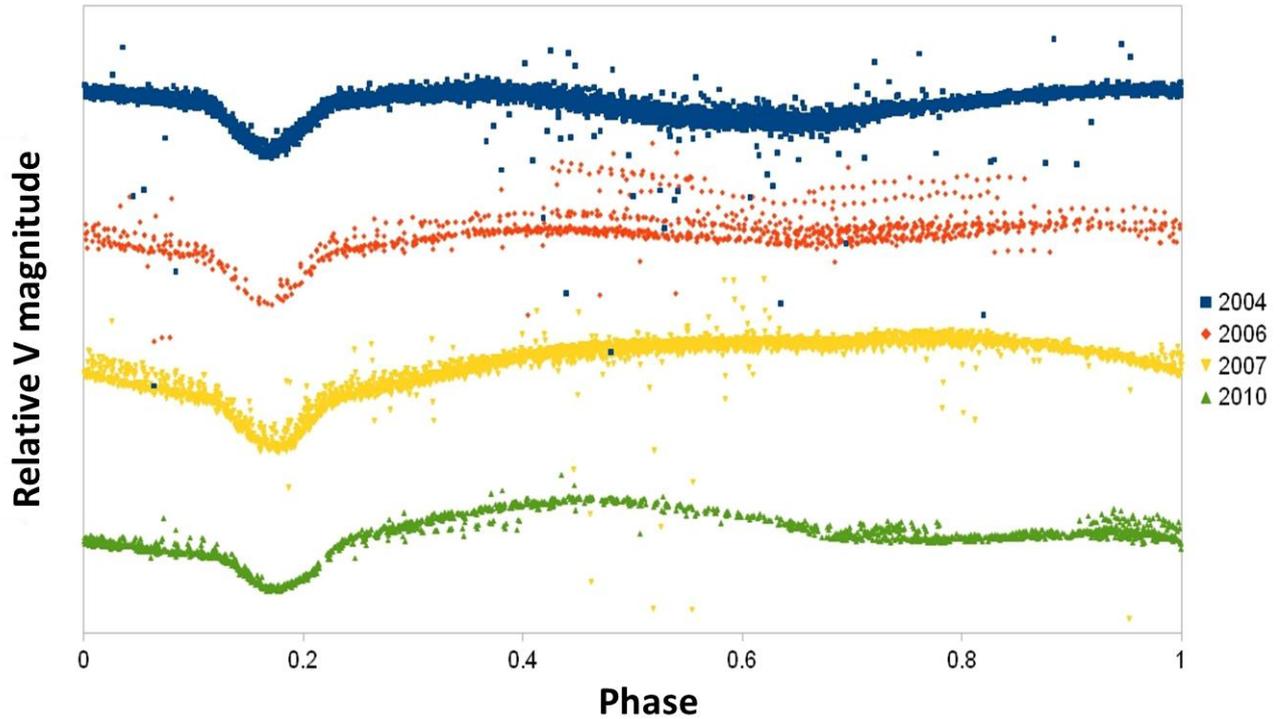

*Figure 8. Our data for 2010 shown against previous GSC 02038-00293 data for different years shifted in y-direction*

There are marked similarities between our data and the 2006 curve by Bernhard and Frank [4] which suggests a shorter cycle of ~6 years duration rather than the previously quoted 6 – 8 year cycle duration.    (Note: The 2004 and 2007 data used only one SWASP survey camera which greatly reduces the problems when comparing magnitudes.   2006 uses primarily two different cameras and also some data from two further cameras.   This explains the increased spread of the data for that year).

When the peak to peak amplitude of the secondary minima are plotted over time, the pattern is more apparent.    Figure 9 shows how the amplitude of the second half of the light curve (phase 0.25 onwards) shows two clear maxima in 1999 and 2005 and minima in 2001-2002 and 2007. 2007 was also the year of lowest starspot activity so far recorded. This suggests that there are cyclical variations possibly in the order of 6 to 8 years. Similar cycles have been observed for other RS CVn stars (Berdyugina and Tuominem 1998) [7].

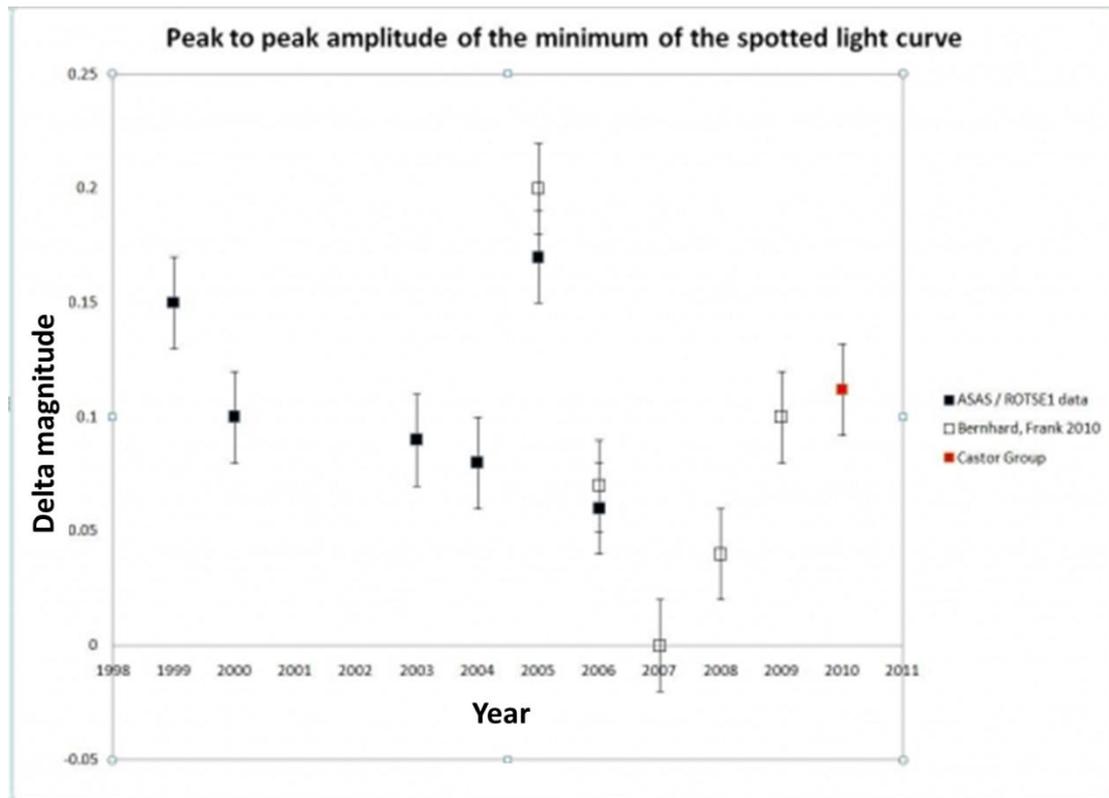

Figure 9: *Peak to peak amplitude of the minimum of the spotted light curve (=amplitude of the "secondary" minimum) in the ASAS and ROTSE1 data (filled squares) Bernhard and Frank V-band data (open squares) in 1999-2007 and our data 2010 (red square)*

Bernhard and Frank (2006 - 2010) [4][6] state that the starspot activity on the larger star is the chief cause of the variation in the depth of the secondary minima. Times of low activity correspond to the data sets where the secondary minima are all but absent. The implications of this are that the secondary eclipse is either very shallow or barely detectable which places restrictions on the size/temperature of the secondary component.

Kozhevnikova, et al. (2007)[8], in a study of 3 RS CVn systems state that star-spotted regions were concentrated at low latitudes up to 32°, that the spots covered up to 29% of the stellar surface and all stars showed non-axisymmetric spot distributions (active longitude structures separated by approximately half the orbital period). This half-phase separation seems to be a common factor in this type of system. The active regions can migrate or change but seem to remain intact over long periods. There is also evidence from these papers of short-term variability on timescales of weeks rather than years.

Our data also shows scatter around the secondary minimum light curve supporting the probability of short term variability and the presence of star spots and hence implying increased coronal activity. As does Korhonen et al (2010) [16] who observed variable line strength in Hα spectroscopy.

The light curve obtained by Norton et al (2007)[1] is similar in structure to our phased light curve but it is clear that their secondary minimum appears to occur earlier in the cycle than in our 2010 data. This may indicate that star spots migrate and, or vary in size and position.

**Anomalies**

The anomalies in the data referred to in the Results section are discussed here.

1. We note a general brightening over part of the light curves of the target over the nights from the 13 to the 25 May which requires further investigation.    This may be due to calibration issues but observations are not conclusive.

2. Figures 2, 3 and 4 above show a marked increase in scatter for V data beginning at phase 0.5, just prior to the secondary minimum.    This appears for data collected on 10 May and 13 June as a collection of data points between phase 0.5 and 0.6.    An increase in the slope of the curve is also apparent in the R and B traces for observations on 13th June, as sharp upturns at 0.6.    Unfortunately the B and R data immediately following this phase was not obtained.    Scatter around the secondary minimum, caused by star spot activity is a known feature of RS CVn binaries.    Our observations may indicate that, as well as varying over several years, there is shorter term variation due to star spot activity.

3. Figures 2 and 4 also show a second area of data points scatter at phase 0.8 visible in V band traces. This clump was only observed on one particular observing night on 25$^{th}$ May, suggesting either a genuine transient event at the binary or local environmental problems at the observatory. We are aware that flares are common in chromospherically active RS CVn stars and a similar increase can be seen in the light curve of figure one of Zhang (2010)[12] at phase 0.35.    However, their flare data profile is much more sharply defined than ours and we have concluded that this is unlikely to be evidence of a flare.

**RS CVn Classification**

Due to the short period of ~0.4955 days deduced from our observations we can be reasonably certain that tidal forces will ensure that the star's rotation will be synchronised with the orbital period. This is confirmed by Korhonen, et al (2010)[16] who deduced the rotation period to be 0.495410 days using low resolution spectroscopy.

The start and end points of the primary eclipse are also clearly defined despite the brightness asymmetry indicating the components are detached.

Dragomir, et al (2007) [11] obtained evidence of CaII, H and K lines with strong core emission which strongly supports RS CVn classification.

**Binary Configuration**

Figure 10 provides a qualitative interpretation of the observed light curve based on our understanding of known RS CVn stars. The assumption is that there is a marked difference in radii (as stated above) in that one star is a giant and the other a dwarf.

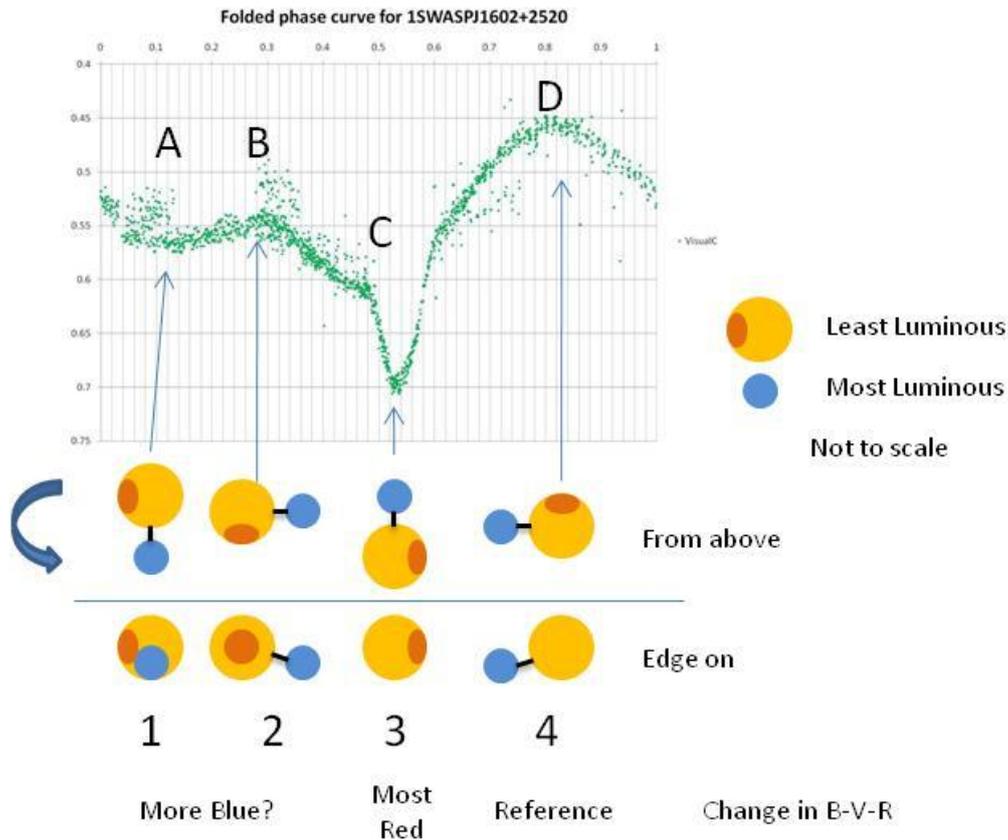

*Figure 10 A speculative model to explain the GSC 02038-00293 light curve.*

The basic configuration shown in figure 10 qualitatively supports the changes in the flux we observed. Point D is the point of greatest luminosity with both stars showing unspotted faces in configuration 4. Point C is the point of lowest luminosity, that being the primary eclipse as shown in configuration 3.

Point B, at configuration 2, is similar to D but with a star spotted primary face which causes a drop in luminosity compared to the reference point D.

Point A should be less luminous than point B but the difference in luminosity will depend on the ratio of the stars' radii, the difference in temperature and the amount of star spotting. If the secondary is a small yellow dwarf with a temperature around 5,000K and since the primary is shown by Korhonen, et al (2010)[16] to be a late K type with a temperature of 4,750K then the radii difference could be > 10:1. This gives a large stellar disc surface area difference and a secondary eclipse may be difficult to ascertain.

It is clear that further modeling, requiring more time and specialist modeling software, should be conducted to establish the actual configuration.

This model also supports the 'reddening' effect that we see in our data and in previous papers.

**Further research**

Further spectroscopy is required to confirm the secondary component (suggested by Korhonen et al (2010) [16]. The determination of the mass ratio and radii would help to refine the starspot models. It would also be interesting to see a 'cleaned' curve that removes the effect of the starspots and to verify any additional cyclic variations in the target.

## 8. Conclusions

We find GSC 02038-00293 is a short period eclipsing RS CVn star. From our data alone we calculate the following ephemeris:

JD 2455327.614 + 0.4955(1) x E

This period is in close agreement with that of Bernhard & Frank (2006)[4] and Norton et al. (2007)[1] but significantly different to that found by the All Sky Automated Survey[15] of 0.330973d.

The eclipse period is very close to the rotation period calculated by Korhonen, et al (2010)[16] of 0.495410 days suggesting a locked synchronous orbit.

Our data shows a clear primary eclipse with notable asymmetry. The secondary minimum is extended and subject to short-term variability. From our observations and the literature, we believe that this is likely to be the result of starspot activity. Due to this we cannot conclusively say the secondary eclipse was detected.

The depth and short-term variability of the secondary minimum is most likely to be a result of starspot activity on one of the system components. This is in line with Bernhard & Frank (2006)[4] and also lends support to their observed 6-8 year cycle of variability in the depth of the secondary minimum. The observed reddening at the minima would also lend support to this hypothesis.

GSC 02038-00293 is a rare, active, very-short period eclipsing RS CVn star which warrants further investigation, specifically the following:

Detailed spectroscopy at other epochs to determine the presence of the binary component.

Detailed light curve modelling allowing subtraction of binary curve effects to allow proper modelling of the spot sizes and locations.

Continuing B, V & R photometric observations to follow secondary changes over the 6-8 year cycle and to determine the nature of the B-V & V-R anti-correlation.

## 9. Acknowledgements

Special thanks to Andrew Norton, Ulrich Kolb, Stefan Holmes and James Smith of the Open University's PIRATE observatory for their guidance and feedback.    Also to Jay Jina, Colm Kilmurry, Brad Rose and Cyrille Tijsseling for their assistance with observing and analysis.

The authors wish to thank the referees whose comments and general guidance greatly improved the paper.## References

1.	Norton A.J., Wheatley P.J., West R.G., Haswell C.A., Street R.A., Collier Cameron A., Christian D.J., Clarkson W.I., Enoch B., Gallaway M., Hellier C., Horne K., Irwin J., Kane S.R., Lister T.A., Nicholas J.P., Parley N., Pollacco D., Ryans R., Skillen I. and Wilson D.M., (2007), 'New periodic variable stars coincident with ROSAT sources discovered using SuperWASP', Astron. Astrophys., 467, 785-905.

2.	J. C. Pandey; K. P. Singh; S. A. Drake; R. Sagar, (2005) "Unravelling the nature of HD 81032 - A new RS CVn binary", Journal of Astrophysics and Astronomy Vol.26 Iss.4 p.359-376.

3.	Hall, D.S., 1976, "The RS CVn Binaries and Binaries with Similar Properties", in Multiple Periodic Variable Stars, (Ed.) Fitch, W.S., Proceedings of IAU Colloquium 29, held in Budapest, Hungary, September 1 – 15, 1975, vol. 60 of Astrophysics and Space Science Library, p. 287, Reidel, Dordrecht; Boston.

4.	Bernhard K. and Frank P., (2006), 'GSC 2038.0293 is a new short-period eclipsing RS CVn variable', IAU Inform. Bull. Var. Stars, 5719, 1-4.

5.	Holmes S, Kolb U, Haswell CA, Burwitz V, Lucas RJ, Rodriguez J, Rolfe SM, Rostron J, Barker J, 2011, "PIRATE: A Remotely-Operable Telescope Facility for Research and Education", PASP, 123, 1177-1187.

6.	Frank P. and Bernhard, K., (2007), 'CCD Photometry of the Short-period Eclipsing RS CVn Variable GSC 2038.0293', Open European Journal on Variable stars, 71, 1.

7.	Berdyugina S. V. (2005)   'Starspots: A Key to the Stellar Dynamo', Living Reviews in Solar Physics 2, 8.

8.	Kozhevnikova, A. V.; Alekseev, I. YU.; Heckert, P. A.; Kozhevnikov, V. P. (2007) "Long-term starspot activity of three short-period RS CVn stars: BH Vir, WY Cnc and CG Cyg" Astronomical & Astrophysical Transactions Vol.26 Iss.1-3 p.111-112.

9.	Bernhard K. and Frank P., (2008) 'Wie lange dauert ein Sternfleckenzyklus auf GSC 2038.0293?', BAV Rundbrief, 57, 163.

10.	Vivekananda Rao (2004).

11.	Dragomir D., Roy P. and Rutledge R.E. [online] (Accessed 27 June 2010), (2007) 'Spectral classification of optical counterparts to ROSAT all-sky survey X-ray sources.' Astron. J., 133, 2495-2501.

12.	Zhang, L; Zhang, X; Zhu, Z. (2010) The CCD photometric study of newly identified RS CVn star DV Piscium.

13.	SuperWASP can be found at http://www.superwasp.org/

14.	ROSAT can be found at http://www.xray.mpe.mpg.de/cgi-bin/rosat/rosat-survey
16

15.     All Sky Automated Survey can be found at http://www.astrouw.edu.pl/asas/?page=main

16.     Korhonen, H; Vida, K; Husarik, M; Mahajan, S; Szcxygiel, D; and Olah, K.    Photometric and spectroscopic observations of three rapidly rotating late-type stars: EY Dra, V374 Peg and GSC 02038-00293.

17     PERANSO light curve and period analysis software can be found at www.peranso.com

18.     Lucas RJ, Kolb U, 2011, "Software architecture for an unattended remotely controlled telescope", Journal of the British Astronomical Association, Volume 121, No 5, 265-269